\documentclass[pdflatex,tex,twocolumn,epjc3]{svjour3}
\RequirePackage[T1]{fontenc}
\usepackage{color}
\newcommand{\beqa}{\begin{eqnarray}}
\newcommand{\eeqa}{\end{eqnarray}}
\usepackage{float}
\usepackage{epsfig}
\usepackage{graphicx}
\usepackage{latexsym}
\usepackage{cite}
\usepackage{mathrsfs}
\usepackage{dcolumn}
\usepackage{bm}%

\usepackage{indentfirst}
\usepackage{amssymb,dcolumn}
\usepackage{latexsym}

\usepackage{setspace}
\usepackage{graphicx,float}
\usepackage[colorlinks,linkcolor=blue,citecolor=blue,urlcolor=blue,hyperindex]{hyperref}
\usepackage{color}
\usepackage{slashed}
\usepackage{amsmath}

\journalname{Eur. Phys. J. C}
\begin{document}

\title{Constraining primordial black holes and primordial curvature power spectrum with extragalactic muon neutrino}

\author{Yupeng Yang\thanksref{addr1,e1}}
\thankstext{e1}{e-mail: ypyang@qfnu.edu.cn}

\institute{School of Physics and Physical Engineering, Qufu Normal University, Qufu, Shandong, 273165, China \label{addr1}}

\date{Received: date / Accepted: date}

\maketitle

\begin{abstract}

We investigate a mixed dark matter scenario comprising weakly interacting massive particles (WIMPs) and primordial black holes (PBHs). After PBH formation, WIMPs can accrete onto them, forming ultracompact minihalos (UCMHs). The resulting WIMP number density within UCMHs is significantly enhanced compared to classical dark matter halo models, 
leading to a higher WIMP annihilation rate. Previous studies have focused mainly 
on the associated gamma-ray flux, we investigate the extragalactic neutrino flux from such annihilation. Considering the annihilation channels $\mu^{+}\mu^{-}$, $\tau^{+}\tau^{-}$, and $\nu_{\mu}\bar{\nu}_{\mu}$, we analyze two classes of neutrino events: upward and contained events. 
By requiring the neutrino flux from WIMP annihilation around PBHs does not exceed the atmospheric neutrino background, we derive upper limits on the fraction of dark matter in PBHs ($f_{\rm PBH}$) for a one-year exposure of the IceCube experiment. These limits depend on the annihilation channel, the masses of the WIMP and PBH, and the neutrino event type. The strongest constraints come from the $\nu_{\mu}\bar{\nu}_{\mu}$ channel, yielding $f_{\rm PBH} \sim 10^{-4}$ ($4\times 10^{-5}$) for contained (upward) events with $m_{\chi}=10^{3}$ GeV and $M_{\rm PBH}=10^{3} M_{\odot}$. Based on these bounds on PBHs, we further derive upper limits on the primordial curvature 
power spectrum $\mathcal{P}_{\mathcal{R}}$. From our strongest constraint, we obtain $\mathcal{P}_{\mathcal{R}} \sim 10^{-1.65}$ 
at the scale $k\sim 3\times 10^{12}~\mathrm{Mpc^{-1}}$. 

\end{abstract}

\maketitle

\section{Introduction}

The existence of dark matter, a major component of the universe, has been firmly established through numerous astronomical observations. 
However, its fundamental nature remains unknown, and a variety of models have been proposed. The most extensively studied candidate is 
weakly interacting massive particles (WIMPs)~\cite{Roszkowski:2004jc,Schumann:2019eaa,2005PhR...405..279B,2021arXiv210905854G,1996PhR...267..195J}. Although many experiments have been conducted to detect WIMPs, 
no confirmed signal has been observed to date~\cite{XENON:2018voc,Arcadi:2017kky,Feng:2022rxt,Schumann:2019eaa}. Other dark matter models, 
such as axion and primordial black holes (PBHs), have also attracted significant research attention 
in recent years~\cite{Carr:2009jm,Choi:2022btl,Nurmi:2021xds}. In particular, the detection of gravitational waves from black hole mergers 
has suggested that some of these signals could originate from the coalescence of PBHs~\cite{PhysRevLett.116.201301,PhysRevLett.120.191102,Hertzberg:2019exb}.

According to theoretical models, PBHs can be formed from the gravitational collapse of 
large density perturbations (e.g., $\delta\rho/\rho \gtrsim 0.3$) in the early universe~\cite{Carr:2009jm,2021RPPh...84k6902C}. 
This distinguishes them from astrophysical black holes, which originate from the collapse of massive stars at the end of their life cycles~\cite{Bambi:2019xzp}. 
The mass of a PBH is related to the cosmic time of its formation via $M_{\mathrm{PBH}} \sim 10^{15} (t / 10^{-23}\,\mathrm{s})\,\mathrm{g}$. 
Consequently, PBH masses span an extremely broad range, from about $10^{-5}\,\mathrm{g}$ at the Planck time ($t \sim 10^{-43}\,\mathrm{s}$) 
up to intermediate-mass scales of order $10^{5}\,M_{\odot}$ for PBHs formed around $t \sim 1\,\mathrm{s}$. 
As shown by Hawking, the temperature of a black hole scales inversely with its mass: $T_{\mathrm{BH}} \sim 10^{-7} (M_{\rm BH}/M_{\odot})^{-1}\,\mathrm{K}$, 
and its lifetime is correspondingly given by $\tau \sim 10^{64} (M_{\rm BH}/M_{\odot})^{3}\,\mathrm{yr}$
~\cite{Carr:2009jm,Hawking:1974rv,Hawking:1974rv,Hawking:1975vcx}. Given the current age of 
the universe ($\sim 13.7$ billion years~\cite{2020A&A...641A...6P}), PBH with masses $M_{\mathrm{PBH}} \lesssim 10^{15}\,\mathrm{g}$ would therefore 
have evaporated completely by the present epoch. PBHs with $M_{\mathrm{PBH}} > 10^{15}~\mathrm{g}$ are still present today and 
could constitute all or part of the dark matter. The fraction of dark matter in the form of PBHs, 
$f_{\mathrm{PBH}} = \rm \Omega_{\mathrm{PBH}}/\Omega_{\mathrm{DM}}$, has been constrained by various observations across different PBH mass ranges~\cite{Carr:2009jm,2021RPPh...84k6902C,Adamek:2019gns,2021PhRvD.103l3532T,Berteaud:2022tws,Xie:2024eug,Bernal:2022swt,Tan:2022lbm,Facchinetti:2022kbg,Coogan:2020tuf,PhysRevD.96.083524,Auffinger:2022khh,PhysRevLett.120.191102,Khlopov_2006,Jung:2019fcs,Mroz:2024wia,Su:2024hrp,Yang:2024pfb,Zhao:2024jad,Cang:2020aoo,Zhang:2023rnp,Yuan:2023bvh,Clark:2018ghm,Huang:2024xap,Boudaud:2018hqb,Yang:2020egn,2021MNRAS.508.5709Y,Yang:2021idt,Yang:2022puh}. 

Although WIMPs have not yet been directly detected, they remain a compelling dark matter candidate. Current research indicates that PBHs 
across a broad mass range cannot account for all of dark matter~\cite{Carr:2009jm}. Consequently, a mixed scenario involving both WIMPs and PBHs has been proposed and explored 
in earlier studies~\cite{Eroshenko:2016yve,2022PhRvD.106d3516Y,2020EPJP..135..690Y,Kadota:2021jhg,2023EPJC...83..934Y,2020PhRvR...2b3204S,2021PhRvD.103l3532T,Adamek:2019gns,Boucenna:2017ghj,Boudaud:2021irr,Gines:2022qzy,Yang:2025mkp,Hao:2024hzu,Chanda:2022hls}. The core concept of this scenario is that after PBHs form, WIMPs can accrete onto them, leading to the formation of ultracompact minihalos (UCMHs). Compared to classical dark matter halos, UCMHs originate much earlier and exhibit a higher WIMP number density than, for example, 
the classical Navarro-Frenk-White (NFW) profile~\cite{Navarro:1996gj,Navarro:1995iw}. Since the annihilation rate of WIMPs scales with the square of their number density, a significantly enhanced annihilation 
rate is expected within UCMHs. This enhancement can leave imprints on various cosmological and astrophysical observables, 
such as the cosmic microwave background (CMB)~\cite{2021PhRvD.103l3532T}, global 21-cm signals~\cite{2020EPJP..135..690Y}, the extragalactic gamma-ray background~\cite{2021MNRAS.506.3648C,Adamek:2019gns,Boucenna:2017ghj}, and neutrino fluxes~\cite{Hao:2024hzu,2021MNRAS.506.3648C}. 
For instance, observational data from these probes can be used to constrain the abundance of dark matter in PBHs. Based on measurements of 
the extragalactic gamma-ray background by the Fermi experiment, the fraction of dark matter in PBHs is constrained to $f_{\rm PBH}\sim 10^{-8} (10^{-10})$ for a WIMP mass $m_{\chi}=10^{3}(10)~\rm GeV$, 
for PBH masses $M_{\rm PBH}>10 M_{\odot}$~\cite{Adamek:2019gns,Boucenna:2017ghj}. In another study, the authors of \cite{Hao:2024hzu} considered 
the extragalactic muon neutrino flux produced by WIMP annihilation in UCMHs (only $\mu^{+}\mu^{-}$ annihilation channel). By comparing this flux with the atmospheric neutrino background (ATM), 
they derived the upper limit of $f_{\rm PBH}\sim 10^{-7}$ 
for $m_{\chi}\sim 10^{3}~\rm GeV$ and $M_{\rm PBH}> 10^{-9} M_{\odot}$. In this work, we will extend such analyses to additional 
annihilation channels ($\mu^{+}\mu^{-}$,$\tau^{+}\tau^{-}$, and $\nu_{\mu}\bar \nu_{\mu}$) and a broader range of PBH masses 
($M_{\rm PBH}>10^{-15}M_{\odot}$). 

The constraints on the fraction of PBHs can be translated into bounds on the primordial curvature power spectrum $\mathcal{P}_\mathcal{R}$~\cite{Josan:2009}. On large scales, $\mathcal{P}_\mathcal{R}$ has been tightly constrained. For instance, on large scales $10^{-4}\lesssim k\lesssim 3~\rm Mpc^{-1}$, 
observations on the CMB, Lyman-$\alpha$, and large scale structure yield a precise estimate of $\mathcal{P}_\mathcal{R}\sim 10^{-8}$~\cite{cmb_2,lyman,large,Ragavendra:2024yfp}. 
On small scales $3\lesssim k\lesssim 10^{23}~\rm Mpc^{-1}$, the available constraints are 
upper limits, reaching $\mathcal{P}_\mathcal{R}\sim 10^{-2}$, which largely comes from PBH bounds~\cite{Josan:2009,PhysRevD.100.063521,Yang:2019bkk}. Previous studies, such as Refs.~\cite{Bringmann:2011ut,Scott:2009tu,yyp_neutrino,Yang_2011,Yang:2011jb}, have examined 
UCMHs formed in the absence of PBHs. Given the properties of such PBH-free UCMHs, their abundance yields an upper 
limit $\mathcal{P}_\mathcal{R}\sim 10^{-7}$ over scales $3\lesssim k\lesssim 10^7~\rm Mpc^{-1}$~\cite{Bringmann:2011ut,yyp_neutrino,Emami_2018}. In this work, based on constraints 
on the PBH fraction derived from the muon neutrino flux, we also derive upper limits on $\mathcal{P}_\mathcal{R}$ over scales range 
$10^{5}\lesssim k\lesssim 10^{14}~\rm Mpc^{-1}$. 
We found that these limits are stronger than those derived from PBHs-only analyses over scales $10^{7}\lesssim k\lesssim 10^{13}~\rm Mpc^{-1}$.

This paper is structured as follows. In Section~\ref{sec:basic}, we outline the basic properties of UCMHs arising from the accretion of WIMPs 
onto PBHs after their formation. Section~\ref{sec2} investigates the extragalactic muon neutrino flux from UCMHs due to WIMP annihilation, 
leading to derived upper limits on the fraction of dark matter in PBHs. The corresponding constraints on the primordial 
curvature power spectrum are presented in Section~\ref{cons}. Finally, the conclusions are given in Section~\ref{con}.


\section{The basic properties of UCMHs }
\label{sec:basic}

In a mixed dark matter scenarios consisting of WIMPs and PBHs, WIMPs can accrete onto PBHs after the latter form, 
leading to the formation of ultracompact minihalos (UCMHs). The density profile of WIMPs, $\rho_{\rm DM}(r)$, within a UCMH varies with radius 
and depends on the masses of the PBH and the WIMP. A detailed discussion of these issues can be found in, e.g., Refs.~\cite{2021MNRAS.506.3648C,Boudaud:2021irr,Boucenna:2017ghj,Eroshenko:2016yve}; here we summarize the main conclusion. 

For relatively massive PBH and WIMP, the density profile scales from $\rho_{\rm DM} \propto r^{-3/2}$ 
in the inner region to $r^{-9/4}$ in the outer region. For lighter PBH and WIMP, it transitions from $r^{-3/4}$ to $r^{-3/2}$ and 
finally to $r^{-9/4}$ from the inside out. This behaviour is rooted in the competition between the WIMP kinetic energy and the gravitational potential energy of 
the PBH, as well as the low‑velocity tail of the WIMP velocity distribution. Specifically, when the kinetic energy is 
negligible compared to the potential energy, the profile follows $r^{-9/4}$. Inside the radius $r_k$ (Eq.~(\ref{eq:r_rc})), 
the kinetic energy becomes dominant, flattening the profile to $r^{-3/2}$. At even smaller radii $r_c$ (Eq.~(\ref{eq:r_rc})), 
the low‑velocity tail of the Maxwellian distribution further softens the profile to $r^{-3/4}$, 
with the density capped by the background value at kinetic decoupling $\rho_{\rm KD}$. 
The specific form of the WIMP density profile in a UCMH is 
given below~\cite{2021MNRAS.506.3648C,Boudaud:2021irr,Boucenna:2017ghj,Eroshenko:2016yve}

\beqa
\rho_{\rm DM}(r)=\left\{
\begin{array}{rcl}
&&f_{\rm DM}\rho_{\rm KD}\left(\frac{r_{c}}{r}\right)^{3/4}, ~~~~~~~~~~~~~~~~~~{r\leq r_{c}}\\
\\
&&f_{\rm DM}\frac{\rho_{\rm eq}}{2}\left(\frac{M_{\rm PBH}}{M_{\odot}}\right)^{3/2}\left(\frac{\hat r}{r}\right)^{3/2},~~{ r_{c}\leq r\leq r_k}\\
\\
&&f_{\rm DM}\frac{\rho_{\rm eq}}{2}\left(\frac{M_{\rm PBH}}{M_{\odot}}\right)^{3/4}\left(\frac{\bar r}{r}\right)^{9/4},~~~~~~~{ r >r_k}
\end{array} \right.
\label{eq:rho_r}
\eeqa
where $f_{\rm DM}\simeq 1$ denotes the fraction of dark matter in the form of WIMPs, 
and $\rho_{\rm eq}$ is the cosmological density at the epoch of radiation-matter equality. The characteristic radii $\hat r$ and $\bar r$ 
are defined as  

\beqa
{\hat r} = \frac{GM_{\odot}}{c^2}\frac{t_{\rm eq}}{t_{\rm KD}}\frac{m_{\rm DM}}{T_{\rm KD}}, ~~~{\bar r}=\left(2GM_{\odot}t^{2}_{\rm eq}\right)^{1/3},
\label{eq:r_ta}
\eeqa
and the transition radii $r_c$ and $r_k$ determined by continuity of the density profile at the junctions between power-law segments, are given by

\beqa
r_{c} = \frac{r_s}{2}\left(\frac{m_{\rm DM}}{T_{\rm KD}}\right), ~~~r_{k}=4\frac{(ct_{\rm KD})^2}{r_S}\left(\frac{T_{\rm KD}}{m_{\rm DM}}\right)^2,
\label{eq:r_rc}
\eeqa
with $r_s=GM/c^2$ being the Schwarzschild radius. The kinetic decoupling time $t_{\rm KD}$ and temperature $T_{\rm KD}$ 
can be expressed as~\cite{Boucenna:2017ghj}, 

\beqa
t_{\rm KD} = \frac{2.4}{g_{\rm KD}}\left(\frac{T_{\rm KD}}{\rm 1MeV}\right)^{-2}s, ~T_{\rm KD}=\frac{m_{\rm DM}}{\Gamma[3/4]}
\left(\frac{{\alpha}m_{\rm DM}}{M_{\rm Pl}}\right)^{1/4},
\label{eq:t_kd}
\eeqa
where ${\alpha}=\sqrt{16\pi^{3}g_{\rm KD}/45}$ with $g_{\rm KD}=61.75$. The cosmological density 
at the time of kinetic decoupling is $\rho_{\rm KD}={\Omega_m}\rho_{\rm eq}(t_{\rm eq}/t_{\rm KD})^{3/2}$~\cite{Eroshenko:2016yve}. 

When WIMPs annihilation is taken into account, the center density of UCMHs is smoothed, and the maximum density at the center 
is~\cite{Yang:2011ef,PhysRevD.72.103517,Hao:2024hzu}, 

\beqa
\rho_{\rm max} = \frac{m_{\chi}}{\left<\sigma v\right>(t(z)-t_{i})},
\label{eq:rho_max}
\eeqa
where $t_{i}$ is the formation time of UCMHs, $m_{\chi}$ is the WIMP mass, and 
$\left<\sigma v\right>$ is the thermally averaged annihilation cross section. In our calculations we adopt the canonical value 
$\left<\sigma v\right> =3\times10^{-26}~\rm cm^3~s^{-1}$. Including the effect of WIMPs annihilation, 
the final density profile of particle dark matter~\footnote{Here we treat WIMPs as particle dark matter.} within a UCMH is taken as  

\beqa
\rho(r,z)={\rm min}\left[\rho_{\rm DM}(r), \rho_{\rm max}\right] 
\label{eq:rho_final}  
\eeqa

\section{Extragalactic muon neutrino flux from WIMP annihilation in UCMHs and constraints on PBHs}
\label{sec2}

\subsection{Muon neutrino flux from UCMHs}

Similar to the extragalactic gamma-ray background from UCMHs due to WIMP annihilation, the neutrino flux can be written as~\cite{Ullio:2002pj,Hao:2024hzu}

\begin{equation}
\begin{split}
\frac{d\phi_{\rm \nu}}{dE_{\rm \nu}}=&\frac{f_{\rm PBH} {\rm \Omega}_{\rm DM}}{M_{\rm PBH}}\frac{c}{8\pi}\frac{\left<\sigma v\right>}{m_{\rm \chi}^2}\\&\times\int_{0}^{z_{\rm up}}\,\frac{dz}{H(z)}\frac{dN_{\rm \nu}}{dE_{\rm \nu}}(E^\prime,z)\int\rho^2(r,z) 4\pi r^{2} dr,
\label{eq:neutrinos_flux}
\end{split}
\end{equation}
where $f_{\rm PBH}=\rm \Omega_{\rm PBH}/\Omega_{\rm DM}$ is the fraction of dark matter in PBHs, 
$E^\prime = E(1 + z)$ and $z_{\rm up} = m_{\chi}/E-1$. $dN_{\rm \nu}/{dE_{\rm \nu}}$ is the energy spectrum of neutrino per WIMP annihilation in UCMHs, 
and it can be obtained using the public code, e.g., $\mathtt{DarkSUSY}$~\cite{Bringmann:2018lay,Gondolo:2004sc}.~\footnote{https://darksusy.hepforge.org/}

Within the Standard Model, neutrinos exist in three flavors: $\nu_e(\bar{\nu}_e)$, $\nu_\tau(\bar{\nu}_\tau)$ and $\nu_\mu(\bar{\nu}_\mu)$. 
These neutrinos undergo vacuum oscillations, leading to flavor transitions during propagation. For simplicity, we assume a uniform flavor ratio of 1:1:1. 
Neutrinos produced by WIMP annihilation in UCMHs propagate to Earth without significant energy attenuation. Upon reaching the Earth, 
muon neutrinos ($\nu_{\mu}$) can undergo charged-current interactions with terrestrial matter (e.g., rock or ice), producing muons ($\mu$). 
These muons are detectable via Cherenkov radiation in subsurface observatories~\cite{Decoene:2023beq,2002ARNPS..52..153G,2009NIMPA.602....7R,2009PhLB..678..101H}. In the context of muon detection, we consider two primary event classes: contained events and upward events~\footnote{Here we refer to ``upward-going events'' simply as ``upward events''.}.

For contained events, muons are generated within the detector volume via charged-current interactions. 
The differential muon flux can be expressed as~\cite{2009PhRvD..80d3514E}

\beqa
\begin{split}
\frac{d\phi_{\rm \mu}}{dE_{\rm \mu}}\bigg|_{\rm con}=&\frac{N_{A}\rho}{2}\int_{E_{\rm \mu}}^{m_{\rm \chi}}\,dE_{\rm \nu}\left(\frac{d\phi_{\rm \nu}}{dE_{\rm \nu}}\right)\\ &\times\left(\frac{d\sigma_{\rm \nu}^p(E_{\rm \nu},E_{\rm \mu})}{dE_{\rm \mu}}+(p\to n)\right)+(\nu \to \bar{\nu}),
\label{eq:flux_ct}
\end{split}
\eeqa 
where $\rho$ is the density of the medium, and $N_{A}=6.022\times10^{23}$ is Avogadro's number. The terms 
$\frac{d\sigma_{\rm \nu,\bar{\nu}}^{p,n}}{dE_{\rm \mu}}$ denote the differential scattering cross sections for neutrinos and antineutrinos 
off protons and neutrons. For these cross sections we adopt the following form as given in Refs.~\cite{2006hep.ph....6054S,2007PhRvD..76i5008B,2009PhRvD..80d3514E}

\beqa
\begin{split}
\frac{d\sigma_{\rm \nu,\bar{\nu}}^{p,n}}{dE_{\rm \mu}}=\frac{2{m_p}{G_F^2}}{\pi}\left(a_{\rm \nu,\bar{\nu}}^{p,n}+b_{\rm \nu,\bar{\nu}}^{p,n}\frac{E_{\mu}^2}{E_{\nu}^2}\right),
\end{split}
\eeqa 
where $a_{\rm \nu}^{\rm p,n}=0.15, 0.25$, $b_{\rm \nu}^{\rm p,n}=0.04, 0.06$ and $a_{\bar{\nu}}^{\rm p,n}=0.06, 0.04, b_{\bar{\nu}}^{\rm p,n}=0.25, 0.15.$ 


For upward events, muons are produced outside the detector as neutrinos arrive from the opposite side of the Earth and traverse its 
interior. The differential muon flux can be written as~\cite{2009PhRvD..80d3514E},

\beqa
\begin{split}
\frac{d\phi_{\rm \mu}}{dE_{\rm \mu}}\bigg|_{\rm up}=&\frac{N_{A}\rho}{2}\int_{E_{\rm \mu}}^{m}\,dE_{\rm \nu} \left(\frac{d\phi_{\rm \nu}}{dE_{\rm \nu}}\right)\\ & \times\left(\frac{d\sigma_{\rm \nu}^p(E_{\rm \nu},E_{\rm \mu})}{dE_{\rm \mu}}+(p\to n)\right)R(E_{\rm \mu})+(\nu \to \bar{\nu}),
\label{eq:flux_up}
\end{split}
\eeqa
where $R(E_{\rm \mu})$ is the distance traveled by the muon in the medium before its energy falls below the detector 
threshold~\cite{2010PhRvD..82b3506Y,2010PhRvD..81h3506S}, and is given by,

\beqa
R(E_{\rm \mu})=\frac{1}{\rho \beta}{\rm ln}\frac{\alpha+\beta E_{\rm \mu}}{\alpha+\beta E_{\rm \mu}^{\rm th}},
\label{eq:rmu}
\eeqa
with $\alpha \sim10^{-3}\, {\rm GeV~{cm}^2~g^{-1}}$ parametrizing ionization energy loss, and $\beta \sim10^{-6}\,{\rm cm}^2~{\rm g}$ representing 
radiative losses from bremsstrahlung and pair production. Here $E_{\rm \mu}^{\rm th}$ is the detector energy threshold, and here we adopt $ E_{\mu}^{th} = 50$ GeV~\cite{2009PhRvD..80d3514E,Hao:2024hzu}.


In neutrino detection, atmospheric neutrinos constitute the dominant background, and it has been detected by related 
experiments~\cite{2002ARNPS..52..153G,2007PhRvD..75d3006H}. For the present analysis, we adopt the following parameterized form of the atmospheric 
neutrino flux~\cite{2007PhRvD..75d3006H},
 
\beqa
\begin{split}
\frac{d\phi_{\rm \nu}}{dE_{\rm \nu}d\Omega}\bigg|_{\rm ATM}=&N_0E_{\rm \nu}^{\rm -\gamma-1} \\
&\times\left(\frac{a}{bE_{\nu}}{\rm ln}(1+bE_{\nu})+\frac{c}{eE_{\rm \nu}}{\rm ln}(1+eE_{\nu})\right),
\end{split}
\eeqa 
where $\gamma=1.74$, $a = 0.018$, $b = 0.024$, $c = 0.0069$, $e = 0.00139$, $N_0 = 1.95 (1.35) \times 10^{17}$ for (anti)neutrinos.

\begin{figure}
\centering
\includegraphics[width=0.5\textwidth]{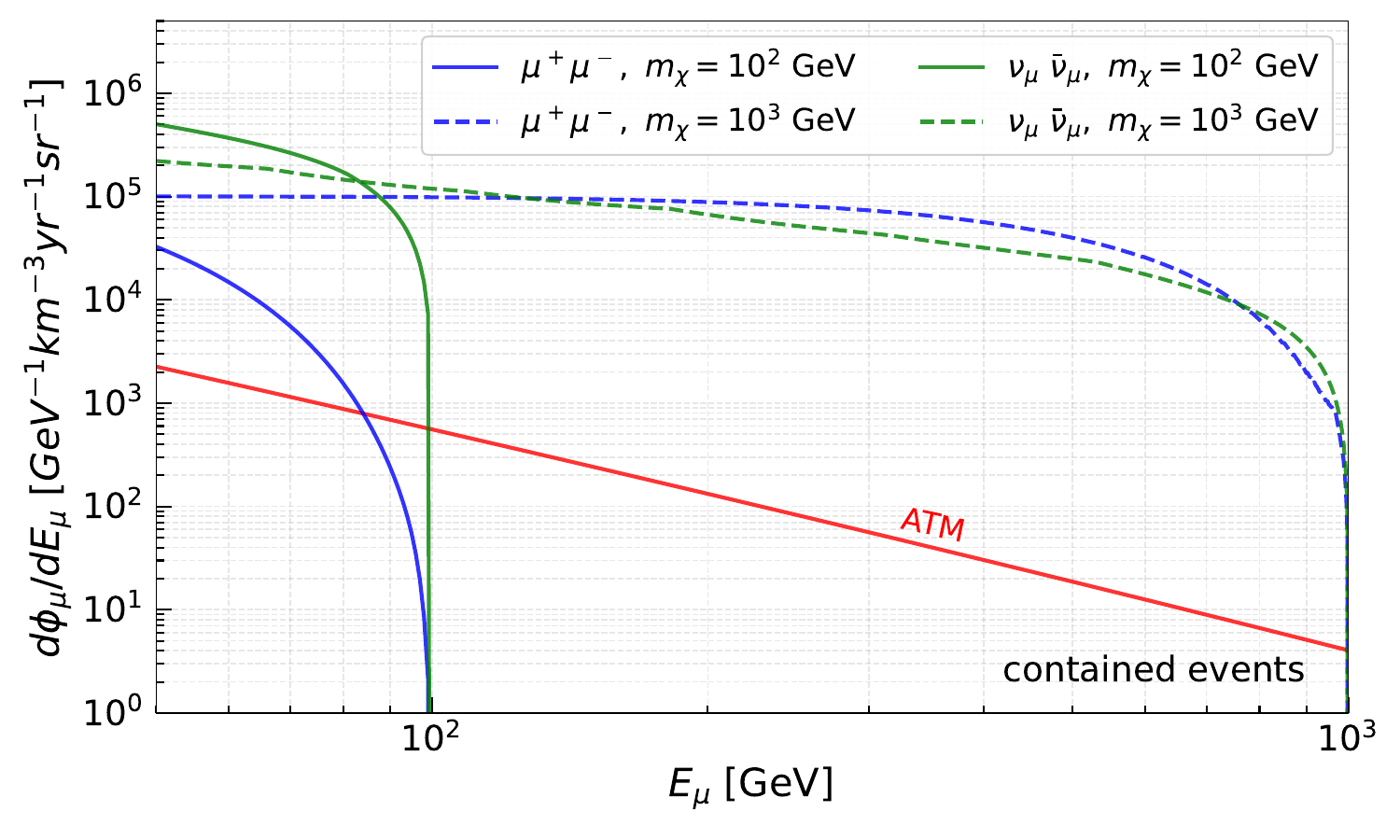}
\includegraphics[width=0.5\textwidth]{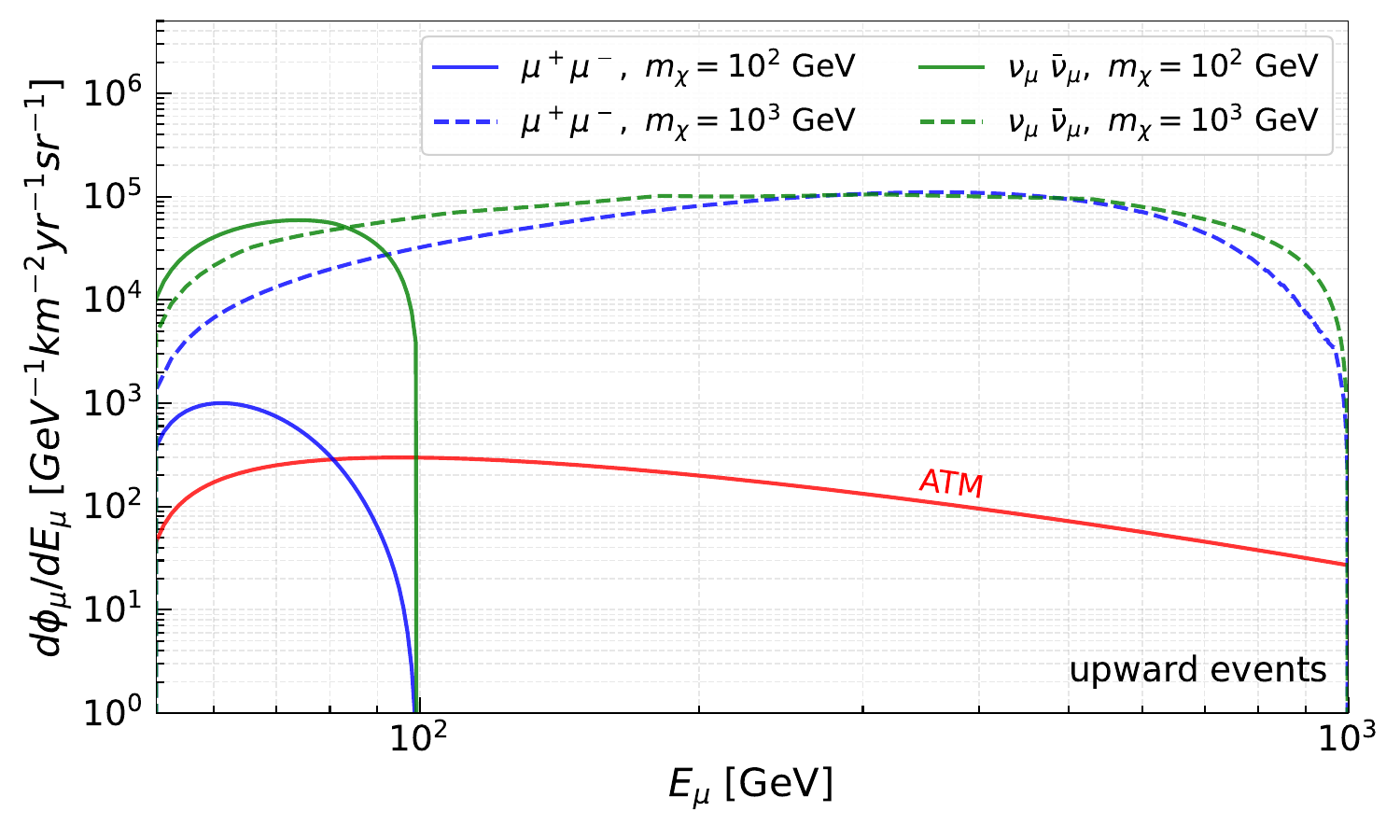}
\caption{Muon flux from WIMP annihilation in UCMHs for $f_{\rm PBH} = 1$. Upper panel: contained events; lower panel: upward events. 
Results are shown for WIMP masses $m_{\chi} = 10^{2}, 10^{3}$ GeV and annihilation channels $\mu^+\mu^-$ and $\nu_{\mu}\bar \nu_{\mu}$ with 
a canonical thermally averaged annihilation cross section $\left<\sigma v\right> =3\times10^{-26}\, \rm cm^3s^{-1}$. 
The atmospheric neutrinos background (ATM) is also plotted for comparison.}
\label{fig:muon_flux}
\end{figure}

Figure~\ref{fig:muon_flux} presents the muon flux for contained (upper panel) and upward (lower panel) events, assuming a PBH fraction $f_{\rm PBH}=1$.
Results are shown for different WIMP masses ($m_{\chi}=10^{2}$ and $10^{3}$ GeV) and annihilation channels ($\mu^{+}\mu^{-}$ and $\nu_{\mu}\bar \nu_{\mu}$). 
For comparison, the atmospheric neutrinos background is also plotted. 
The flux from the $\tau^{+}\tau^{-}$ channel is comparable to that of $\mu^{+}\mu^{-}$, and is therefore omitted from the plot. 
Depending on the PBHs fraction, the muon flux from UCMHs due to WIMP annihilation can exceed the ATM, 
especially at higher energies (corresponding larger WIMP masses) where the ATM drops significantly relative to the lower energy regime.


\subsection{Constraints on the fraction of dark matter in PBHs}

The number of muon neutrinos ($N_{\nu_\mu}$) originating from extragalactic UCMHs associated with PBHs, 
produced via WIMP annihilation, can be expressed as

\beqa
\begin{split}
N_{\nu_\mu} = \int_{E_\mu^{\mathrm{th}}}^{E_{\max}} \frac{d\phi_\mu}{dE_\mu} F_{\mathrm{eff}}(E_\mu) dE_\mu ,
\label{eq:N_PBHs}
\end{split}
\eeqa
where $d\phi_\mu/dE_\mu$ is given by Eqs.~(\ref{eq:flux_ct}) and (\ref{eq:flux_up}). The factor $F_{\mathrm{eff}}(E_\mu)$ corresponds 
to the effective volume $V_{\mathrm{eff}}$ for contained events, or to the effective area $A_{\mathrm{eff}}$ for upward events. 
In general, both $V_{\mathrm{eff}}$ and $A_{\mathrm{eff}}$ depend on the energy of the detected particles. 
For simplicity, we adopt energy-independent values of, e.g., the IceCube experiment: 
an effective volume $V_{\mathrm{eff}} = 0.04~\mathrm{km}^3$ and an angle-averaged muon effective area 
$A_{\mathrm{eff}} = 1~\mathrm{km}^2$~\cite{2009arXiv0907.2263W,2010PhRvD..81d3508M,2009NIMPA.602....7R}.

To derive constraints on the fraction of dark matter in PBHs $f_{\rm PBH}=\rm \Omega_{\rm PBH}/\Omega_{\rm DM}$, we treat atmospheric neutrinos 
as the dominant background and assume an exposure time of one year. 
Accounting for this background, upper limits on the PBH abundance at, e.g., $2\sigma$ statistical significance can be obtained using 
the statistic~\cite{Fornengo:2011em,Bergstrom:1997tp}

\beqa
\begin{split}
\zeta \equiv \frac{N_{\mathrm{PBH}}}{\sqrt{N_{\mathrm{PBH}} + N_{\mathrm{ATM}}}},
\end{split}
\eeqa
where $N_{\mathrm{ATM}}$ denotes the number of muon neutrinos from atmospheric background.

\begin{figure}
\centering
\includegraphics[width=0.5\textwidth]{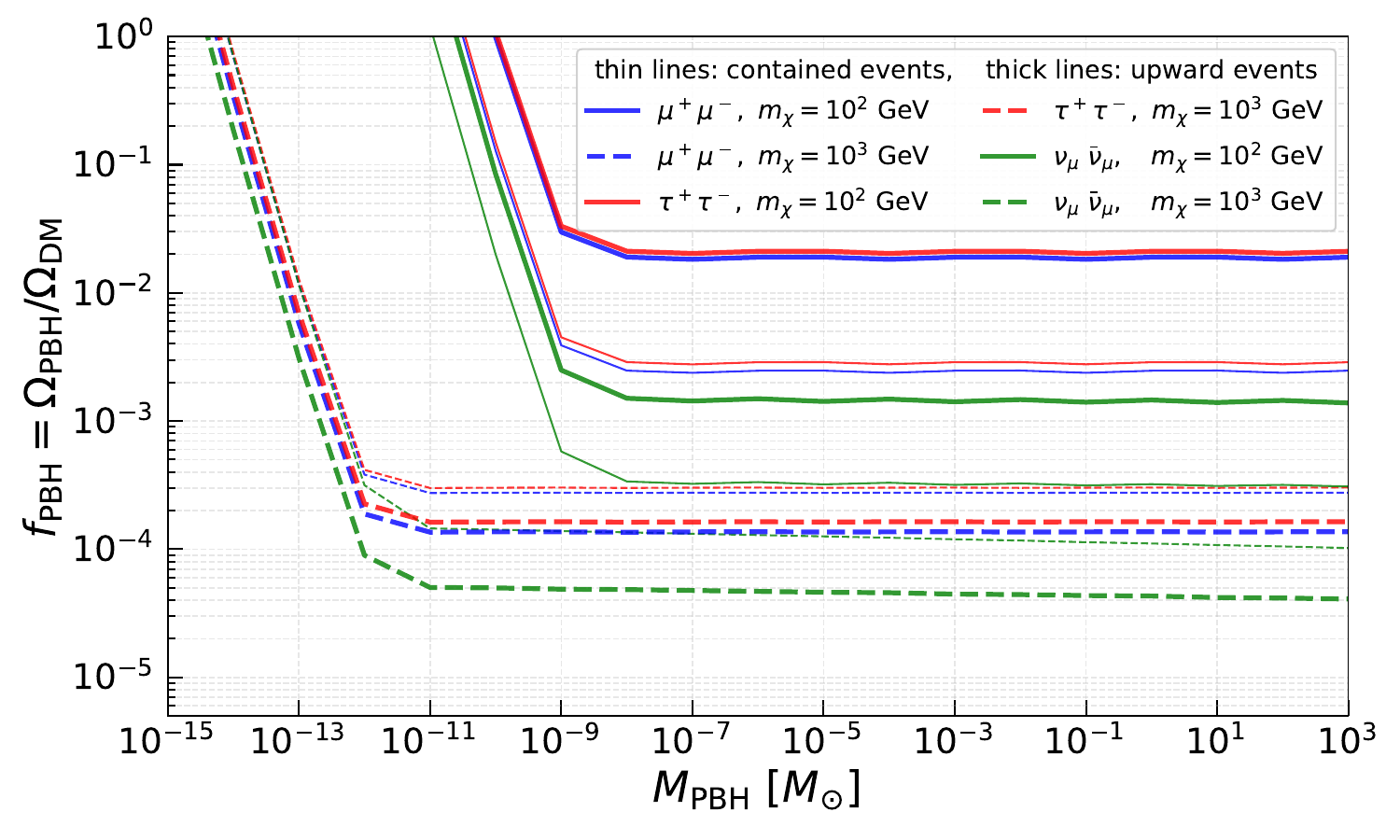}
\includegraphics[width=0.5\textwidth]{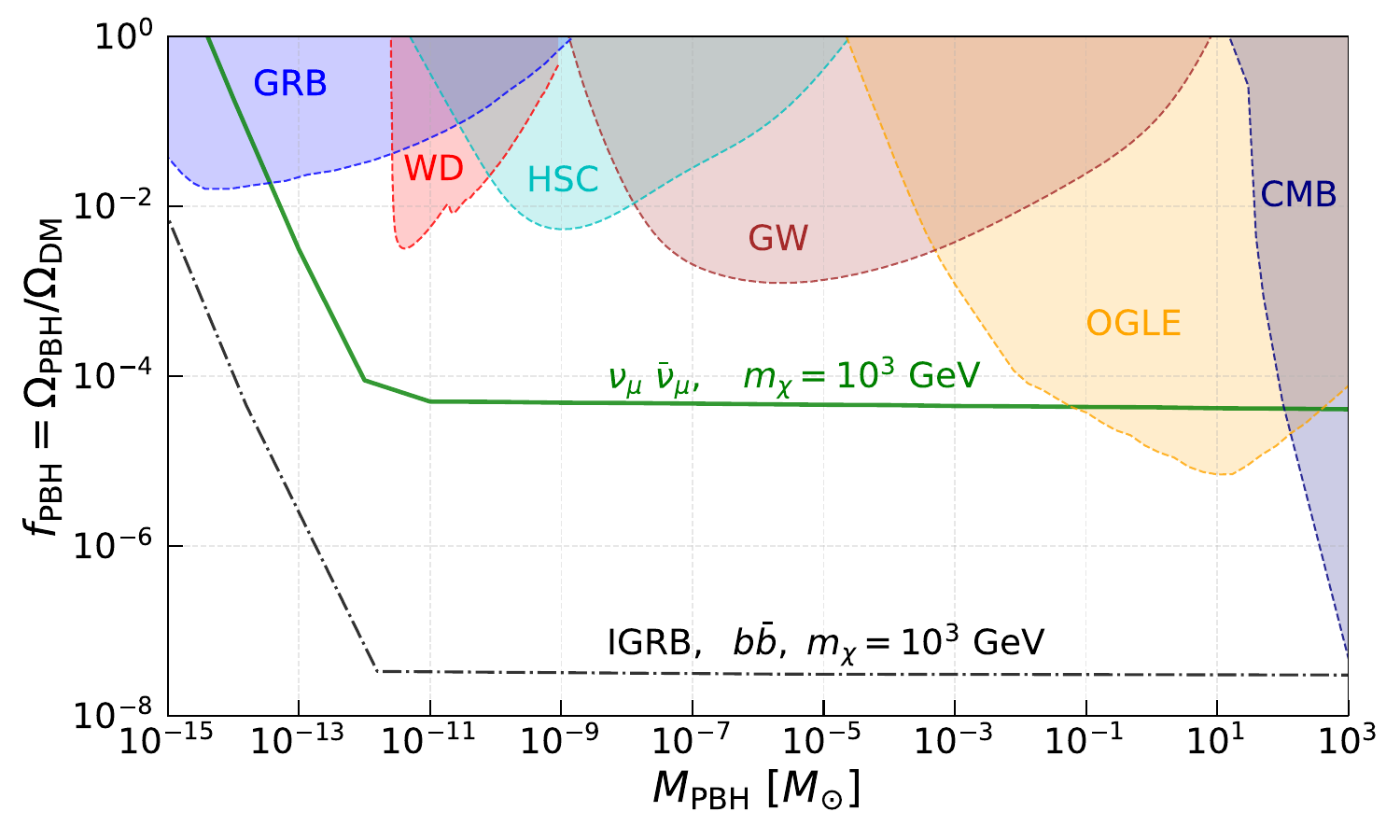}
\caption{Upper limits on the fraction of dark matter in PBHs, $f_{\rm PBH}={\rm \Omega}_{\rm PBH}/{\rm \Omega}_{\rm DM}$, 
are shown in the upper panel for 
annihilation channels $\mu^{+}\mu^{-}$, $\tau^{+}\tau^{-}$, and $\nu_{\mu}\bar \nu_{\mu}$. Constraints are presented for 
WIMP masses $m_{\chi}=10^{2}$ GeV (solid lines) and $10^{3}$ GeV (dashed lines), with contained events indicated by thin lines 
and upward events by thick lines. 
Constraints from other observations are also plotted for comparisons (lower panel):
(1) GRB lensing parallax (GRB)~\cite{Jung:2019fcs}; 
(2) white Dwarf explosions (WD)~\cite{Graham:2015apa}; 
(3) gravitational microlensing from HSC (HSC)~\cite{Croon:2020ouk}; 
(4) 3G gravitational wave detectors with 10 year of observations (GW)~\cite{Kalogera:2021bya}; 
(5) gravitational microlensing events from the OGLE high-cadence survey of the Magellanic Clouds (OGLE)~\cite{Mroz:2024wia}; 
(6) cosmic microwave background affected by accretion radiation from massive PBHs (CMB)~\cite{Serpico:2020ehh}. 
Data are taken from the zenodo~\cite{bradley_j_kavanagh_2019_3538999}. 
Constraints from the isotropic gamma-ray background (IGRB) 
are also shown for the $b\bar b$ annihilation channel (dot-dashed black line)~\cite{Gines:2022qzy}. 
} 
\label{fig:fraction_compare}
\end{figure}

The final constraints on the fraction of dark matter in PBHs are presented in Fig.~\ref{fig:fraction_compare}. 
Results are shown for three annihilation channels: $\mu^{+}\mu^{-}$, $\tau^{+}\tau^{-}$, and $\nu_{\mu}\bar \nu_{\mu}$, for 
the WIMP masses $m_{\chi}=10^{2}$ GeV and $10^{3}$ GeV. For contained events (thin lines in Fig.~\ref{fig:fraction_compare}), 
the constraints from the $\mu^{+}\mu^{-}$ and $\tau^{+}\tau^{-}$ are comparable, with the $\mu^{+}\mu^{-}$ 
slightly stronger. The strongest constraints come from the $\nu_{\mu}\bar \nu_{\mu}$ channel, depending on the WIMP mass. 
The tightest upper limit for contained events is $f_{\rm PBH}\sim 10^{-4}$, obtained for the $\nu_{\mu}\bar \nu_{\mu}$ channel 
with $m_{\chi}=10^{3}$ GeV and $M_{\rm PBH}=10^{3} M_{\odot}$. 
A similar trend is observed for upward events (thick lines in Fig.~\ref{fig:fraction_compare}). Because upward muons travel a longer distance in 
the medium before their energy falls below the detector, a distance that increases with higher initial energy, 
the constraints for upward events are stronger than for 
contained events. The most stringent limit in this case is $f_{\rm PBH}\sim 4\times 10^{-5}$, again for the 
$\nu_{\mu}\bar \nu_{\mu}$ channel with $m_{\chi}=10^{3}$ GeV and $M_{\rm PBH}=10^{3} M_{\odot}$.

The constraints vary with PBH mass depending on the relative importance of the WIMP kinetic energy compared 
to the gravitational potential energy around a PBH. The behavior is divided into two regimes by a characteristic PBH mass scale:
$M_{\rm PBH}\sim 10^{-2}M_{\odot}\left(m_{\chi}/{\rm GeV}\right)^{-73/24}$
~\cite{2021PhRvD.103l3532T}. 
Below this mass, where kinetic energy is non-negligible relative to the potential energy, the upper limit on $f_{\rm PBH}$ 
changes with PBH mass. Above this scale, where kinetic energy is negligible, the limit remains nearly constant. 
Specifically, for $m_{\chi}=10^{2}$ ($10^{3}$) GeV, the upper limits are nearly constant 
for PBH masses above $M_{\rm PBH}\sim 10^{-8}M_{\odot}$ ($10^{-11}M_{\odot}$), consistent with previous works~\cite{2021PhRvD.103l3532T,Hao:2024hzu,2021MNRAS.506.3648C}. 
For lighter PBHs below these masses, the upper limits become progressively weaker, reaching 
$f_{\rm PBH}\sim 1$ for certain combinations of WIMP and PBH masses. 
We also note that the stronger limit comes from contained events for mass $m_{\chi}=10^{2}$ GeV, 
but from upward events for $m_{\chi}=10^{3}$ GeV. This transition is due to the higher-energy muons produced 
in UCMHs for larger WIMP masses, which increases the effective propagation distance $R(E_{\mu})$ in Eq.~(\ref{eq:rmu}).


\section{Constraints on primordial curvature power spectrum}
\label{cons}

Since the formation of PBHs is linked to the density perturbations of early universe, constraints on the 
PBH abundance can be translated into limits on the amplitude of the primordial curvature power spectrum on 
the corresponding scales. For Gaussian initial density distributions~\footnote{Note that the primordial density perturbation could be non-Gaussian, see, e.g., Refs.~\cite{Achucarro:2022qrl,Matsubara:2022nbr,Pi:2024lsu}. The present analysis does not consider this case.}, the probability distribution of 
the smoothed density contrast $\delta (R)$ is given by~\cite{PhysRevD.100.063521,Josan:2009}

\beqa
P(\delta(R))=\frac{1}{\sqrt{2\pi}\sigma(R)}{\rm exp}\left(-\frac{\delta^{2}(R)}{2\sigma^{2}(R)}\right),
\eeqa 
where the mass variance $\sigma^{2}(R)$ takes the form 

\beqa
\sigma^{2}(R)=\int^{\infty}_{0} W^{2}(kR)\mathcal{P}_{\delta}(k,t)\frac{dk}{k} 
\eeqa 
with a window function $W(kR)={\rm exp}(-k^{2}R^{2}/2)$. Here $\mathcal{P}_{\delta}$ denotes the primordial 
density perturbation power spectrum. Within the Press-Schechter formalism, the initial mass fraction of PBHs can be expressed as~\cite{Carr:2009jm} 

\beqa
\beta(M_{\rm PBH})=&&2\int^{1}_{\delta_c}P(\delta(R))d\delta(R)	\nonumber \\
=&&\frac{2}{\sqrt{2\pi}\sigma(R)}\int^{1}_{\delta_c}{\rm exp}\left(-\frac{\delta^{2}(R)}{2\sigma^{2}(R)}\right)d\delta(R)\nonumber \\
\simeq &&{\rm erfc}\left(\frac{\delta_c}{\sqrt{2}\sigma(R)}\right),
\eeqa 
where $\delta_c$ is the critical density contrast required for PBHs formation. This threshold depends on several physical factors, 
and a simple estimate yields $\delta_{c}\sim 0.3$~\cite{Carr:2009jm}, corresponding to the value of equation of state of matter 
at the radiation-dominated epoch. 
More detailed numerical simulations indicate a range $0.42\lesssim \delta_{c}\lesssim 0.66$~\cite{Polnarev:2006aa}, 
depending on the curvature profile.  

The primordial density perturbation power spectrum $\mathcal{P}_{\delta}$ is related to the primordial curvature 
power spectrum $\mathcal{P}_{\mathcal{R}}$ as~\cite{Josan:2009,Bringmann:2011ut,PhysRevD.100.063521}

\beqa
\mathcal{P}_{\delta}=\frac{4(1+w)^{2}}{(5+3w)^{2}}\left(\frac{k}{aH}\right)^{4}\mathcal{P}_{\mathcal{R}},
\eeqa
where $w$ is $1/3$ in the radiation-dominated epoch. 

The initial mass fraction $\beta(M_{\rm PBH})$ 
can be connected to the present abundance $f_{\rm PBH}$ as~\cite{Josan:2009,Carr:2009jm}

\beqa
\beta(M_{\rm PBH})=1.6\times 10^{-25}f_{\rm PBH}\left(M_{\rm PBH}/{\rm g}\right)^{1/2},
\label{eq:initial_f}
\eeqa 
where $f_{\rm PBH} = \rm \Omega_{\rm PBH}/\Omega_{\rm DM}$. For deriving Eq.~(\ref{eq:initial_f}), we have adopted 
a collapse fraction $\gamma=0.2$ of the horizon mass collapses to form PBHs, and taken 
the number of relativistic degrees of freedom at formation as $\rm g_{\star i} \approx 100$~\cite{Carr:2009jm}. 

\begin{figure}
\centering
\includegraphics[width=0.5\textwidth]{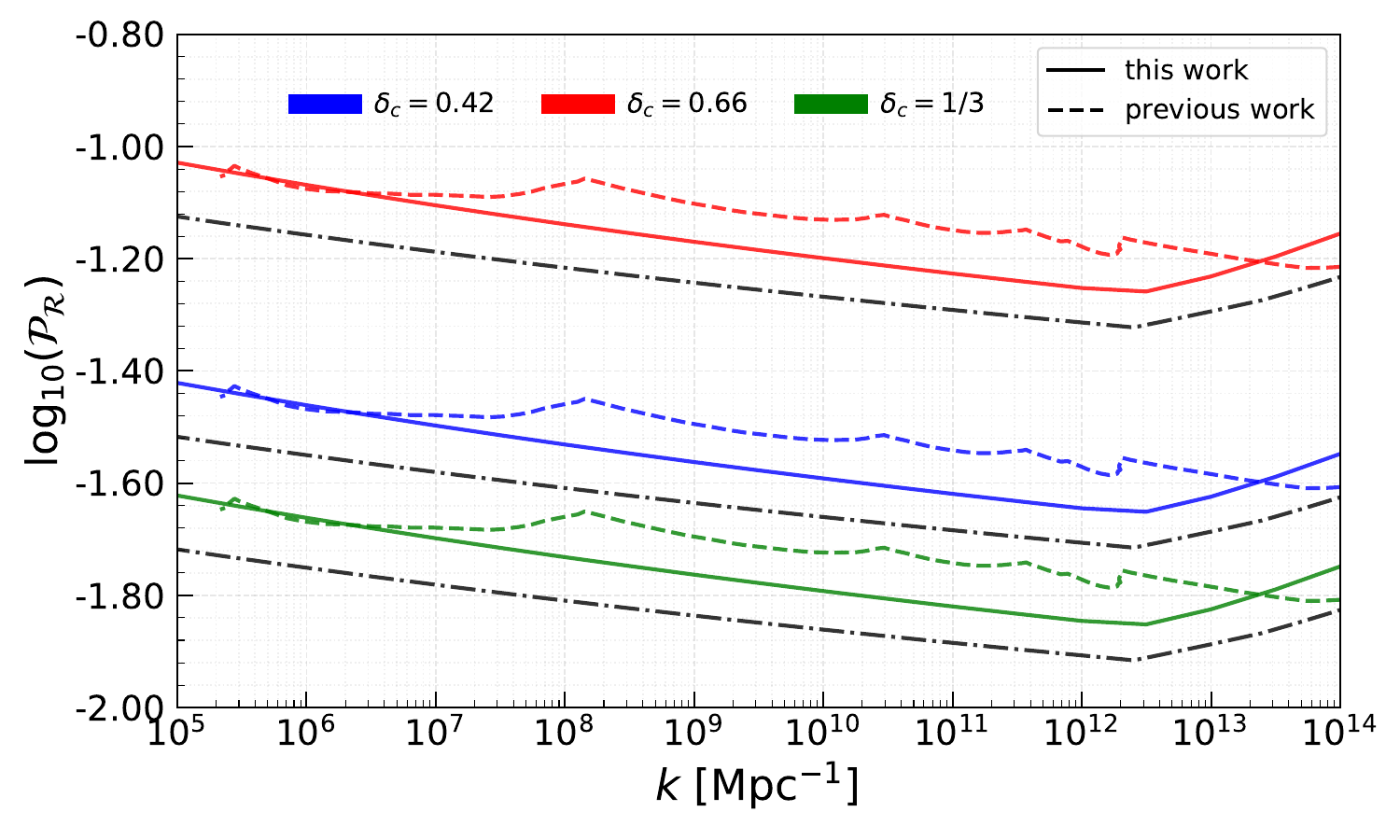}
\caption{Upper limits on the amplitude of the primordial curvature power spectrum $\mathcal{P}_{\mathcal{R}}$ for 
the critical density contrast $\delta_c$ = 1/3 (green), 0.42 (blue), and 0.66 (red). 
The solid lines represent the strongest constraints derived in this work, based on the $\nu_{\mu}\bar \nu_{\mu}$ channel, 
a dark matter mass $m_{\chi}=10^{3}$ GeV, and upward events. The dashed lines show the corresponding combined constraints 
from previous studies, which are displayed in Fig.~\ref{fig:fraction_compare} for comparison. The dot‑dashed black lines 
indicate the IGRB constraints, which correspond to the limits presented in Fig.~\ref{fig:fraction_compare}.
}
\label{fig:pps_compare}
\end{figure}

By applying the previously obtained constraints on PBHs, we derive upper limits on the amplitude 
of the primordial curvature power spectrum $\mathcal{P}_{\mathcal{R}}$. 
The resulting constraints on $\mathcal{P}_{\mathcal{R}}$ are presented in Fig.~\ref{fig:pps_compare} over relevant 
scales range $10^{5}\lesssim k\lesssim 10^{14}~\rm Mpc^{-1}$. It should be noted that these bounds correspond 
to the most stringent limits on $f_{\rm PBH}$ from the $\nu_{\mu}\bar \nu_{\mu}$ channel, for a dark matter mass 
$m_{\chi}=10^{3}$ GeV and considering upward events. For the critical density contrast $\delta_c$ = 0.42, 
the tightest constraint reaches $\mathcal{P}_{\mathcal{R}}\sim 10^{-1.65}$ at the scale $k\sim 3\times 10^{12}~\rm Mpc^{-1}$. 
For $\delta_c$ = 0.66, corresponding to the upper value from simulations, the limits on $\mathcal{P}_{\mathcal{R}}$ are reduced 
by a factor of about 2.3. The constraints for the commonly estimated value $\delta_c$=1/3 are also shown for comparison. 
These limits are strongest, being improved by a factor of about 3.7 compared to the case with $\delta_c$=0.42.
For comparison, constraints from previous works, which are shown in Fig.~\ref{fig:fraction_compare}, are also included. 
The limits on $\mathcal{P}_{\mathcal{R}}$ obtained in this work are stronger than those from previous studies over the scale range 
$10^{7}\lesssim k\lesssim 10^{13}~\rm Mpc^{-1}$. 
Moreover, the IGRB constraints, corresponding to the limits shown in Fig.~\ref{fig:fraction_compare}, 
are also presented. As can be seen in Fig.~\ref{fig:fraction_compare}, the IGRB limits on $f_{\rm PBH}$ 
are about four orders of magnitude stronger than the neutrino limits obtained in this work. 
However, the corresponding IGRB constraints on $\mathcal{P}_{\mathcal{R}}$ are only a factor of about 1.2 
stronger than those from neutrino.

\section{Conclusions}
\label{con}

We have investigated a mixed dark matter scenario consisting of WIMPs and PBHs. After PBHs form, WIMPs can accrete onto them, 
forming ultracompact minihalos. Within such halos, the WIMP number density is significantly enhanced relative to those in classical dark matter halo models, leading to a large increase in the WIMP annihilation rate. While earlier studies have focused mainly on the associated gamma-ray flux, we have examined the extragalactic neutrino flux from WIMP annihilation around PBHs. We considered three annihilation channels: 
$\mu^{+}\mu^{-}$, $\tau^{+}\tau^{-}$, and $\nu_{\mu}\bar \nu_{\mu}$, and analyzed two types of neutrino events: contained events and upward events. 

By requiring that the neutrino flux from WIMP annihilation around PBHs does not exceed the atmospheric neutrino background, 
we derived upper limits on the fraction of dark matter in PBHs, 
$f_{\rm PBH}=\rm \Omega_{\rm PBH}/\Omega_{\rm DM}$, for one-year exposure of experiments such as IceCube. 
The limits depend on the annihilation channels, the masses of WIMP and PBH, and the type of neutrino event. 
In general, for lighter PBHs the constraints weaken with decreasing PBH mass, while for more massive PBHs they 
remain approximately constant. Among the cases studied, the strongest upper limit come from the $\nu_{\mu}\bar \nu_{\mu}$ channel, 
yielding $f_{\rm PBH}\sim 10^{-4}$ for contained events and $\sim 4\times 10^{-5}$ for 
upward events, for $m_{\chi}=10^{3}$ GeV and $M_{\rm PBH}=10^{3} M_{\odot}$. 
Compared with existing bounds from other observations, our results provide valuable complementary constraints on PBH masses 
$M_{\rm PBH} >10^{-15}M_{\odot}$.

Using the obtained limits on the PBH fraction, we further derived upper bounds on the primordial curvature power spectrum 
$\mathcal{P}_{\mathcal{R}}$. From our strongest constraint, corresponding to the $\nu_{\mu}\bar \nu_{\mu}$ channel with $m_{\chi}=10^{3}$ GeV, 
we obtained the strongest upper limit of $\mathcal{P}_{\mathcal{R}}\sim 10^{-1.65}$ at the scale $k\sim 3\times 10^{12}~\rm Mpc^{-1}$. 
Over the corresponding PBH mass range, and compared with previous limits derived from PBHs fraction constraints without considering WIMPs, 
our upper limits on $\mathcal{P}_{\mathcal{R}}$ are stronger over the scale range $10^{7}\lesssim k\lesssim 10^{13}~\rm Mpc^{-1}$.

\section*{Acknowledgements}
This work is supported by the Shandong Provincial Natural Science Foundation (Grant Nos. ZR2025MS16). 

\newcommand{\bibcommenthead}{}
\bibliographystyle{sn-aps}
\bibliography{ref}

\end{document}